\documentclass[10pt,letterpaper]{article}
\usepackage[top=0.85in,left=2.75in,footskip=0.75in]{geometry}

\usepackage{amsmath,amssymb}

\usepackage{changepage}

\usepackage[utf8x]{inputenc}

\usepackage{textcomp,marvosym}

\usepackage{cite}

\usepackage{nameref,hyperref}

\usepackage{microtype}
\DisableLigatures[f]{encoding = *, family = * }

\usepackage[table]{xcolor}

\usepackage{array}

\newcolumntype{+}{!{\vrule width 2pt}}

\newlength\savedwidth

\raggedright
\usepackage[aboveskip=1pt,labelfont=bf,labelsep=period,justification=raggedright,singlelinecheck=off]{caption}
\renewcommand{\figurename}{Fig}

\makeatletter
\renewcommand{\@biblabel}[1]{\quad#1.}
\makeatother

\date{}

\begin{document}
\vspace*{0.2in}

\begin{flushleft}
{\Large
\textbf\newline{The efficiency of community detection by most similar node pairs} 
}
\newline
\\
YunFeng Chang\textsuperscript{1*},
JiHui Han\textsuperscript{2},
\\
\bigskip
\textbf{1} College of Science, China Three Gorges University, YiChang, HuBei, China
\\
\textbf{2} School of Computer and Communication Engineering, Zhengzhou University of Light Industry, ZhengZhou, HeNan, China
\\
\bigskip

%
%





* changyf@ctgu.edu.cn

\end{flushleft}
\section*{Abstract}
Community analysis is an important way to ascertain whether or not a complex system consists of sub-structures with different properties. Avoiding the shortages of computation complexity and pre-given assumption, in this paper, we give a two level community structure analysis for the SSCI journal system by most similar node pairs.  Five different strategies for the selection of node pairs are introduced. The efficiency is checked by normalized mutual information technique. Statistical properties and comparisons of the community results show that both of the two level detection could give instructional information of the community structure of complex systems. Further comparisons of the five strategies indicates that, it is always efficient to assign nodes with maximum similarity into the same community whether the similarity information is complete or not, while rational random selection with too much information and random selection generate small world local community with no inside order. These results give valuable indication for efficient community detection by most similar node pairs.



\section*{Introduction}
Communities are supposed to play special roles in the structure-function relationship. Individuals that share common properties self-organize into communities. For examples, the communities in WWW are sets of web pages sharing the same topic~\cite{1}; the modular structure in biological networks are widely believed to play important roles in biological functions~\cite{2,3,4,5}. The identification of community structure helps when analyzing the functionalities and organizations of complex systems.

Complex network has attracted considerable attention in physics and other fields as a foundation for the mathematical representation of a variety of complex systems, such as biology~\cite{6}, sociology~\cite{7}, medicine~\cite{8}, web~\cite{9}, and many others~\cite{10}. In the field of complex network study, communities are defined as groups of nodes that are densely interconnected but only sparely connected with the rest of the network~\cite{2,11,12,13}. With this network based definition, researchers have proposed different algorithms for the detection of communities, such as topology based methods~\cite{2,14}, modularity optimization~\cite{15,16}, dynamic label propagation~\cite{17,18,19}, statistical inference~\cite{20,21,22}.

Community detection is also called cluster analysis which is done with different kinds of relationships. Specifically, cluster analysis is the assignment of a set of observations into clusters of components that are similar to each other but different from components in other clusters. It is often used to ascertain whether a complex system comprises a set of distinct clusters, each representing components with substantially different properties. On the basis of cluster analysis, some emerging community detection methods are also proposed~\cite{25,26,27,28}.

Community detection by most similar node pairs has been proposed as an efficient method for community analysis~\cite{29,30}. In this paper, we give a two level community structure analysis for the SSCI journal system with five strategies for the selection of most similar node pairs. Emerging characteristic numbers of core-community and real-community correspond to two community detection level: diverse core level and small world real level, which provide different resolution scales for viewing the system and might be helpful in understanding the mutual interactions among various knowledge fields. The efficiency is further checked by normalized mutual information technique. Statistical properties and comparisons of the community results show that both the core and real community could give valuable information for the community structure of complex systems. Further comparisons of the five strategies indicates that it is always efficient to assign nodes with maximum similarity into the same community.

\section*{The method}

In real-life systems, communities are constructed by individuals with the choosing of friends. And this choosing process are based on individual's judgment of its relationship with the surroundings. Most probably, individual chooses the one who is most similar to it or the one satisfy its expectation mostly. So that it might be a good choice to reconstruct and detect communities by the most similar pairs. That means, detecting communities by formalizing those relationships or those components believed to be the most significant. And similarity can be different kinds of interaction according to the properties of complex systems. For examples, internet users with common interests~\cite{32}, social communities with distinctive social norms~\cite{33}, related proteins to execute specific functions~\cite{34}.

The property of a scientific journal system can be well described by the citation pattern of its articles. Journals in the same research field usually have similar citation patterns, while journals in different research fields could have very different citation patterns. The citation pattern of a journal is described by a citation frequency vector $v_{i}$, where $i$ runs over all journals in consideration. The collection of all citation pattern vectors forms a citation matrix $\{N_{ij}\}$. By this citation pattern matrix, we calculate the similarity of two journals $i$ and $j$ in the cosine measure:

\begin{eqnarray}
\label{eq:schemeP}
S_{ij}=\frac{v_{i}\cdot v_{j}}{\|v_{i}\|\|v_{j}\|}=\frac{\sum\limits_{k\in\Omega}(c_{ik}c_{jk})}{\sqrt{\sum\limits_{k\in\Omega}(c_{ik}^2)}\cdot\sqrt{\sum\limits_{k\in\Omega}(c_{jk}^2)}}
\end{eqnarray}

Where $c_{ik}=N_{ik}/(\sum\limits_{j\in\Omega}N_{ij})$ is the normalized citation matrix element, and the value of $S_{ij}$ ranges from 0 to 1. Such that closely related journals have strong similarity while remotely related journals have small similarity.

The detection process is carried out in the following steps:

 {\bf Step 1: Selection of node pairs.} Find the journal or journals with maximum similarity for each journal and record their similarity in a decreasing list (MAX strategy). If there is more than one journal that has the same maximum similarity with the same journal, all these most similar journals will be assigned into the same community. This step results in most similar node pairs of size $O(N)$ ($N$ is the total number of journals), which reduces the computation complexity from $O(N^{2})$  to $O(N)$.

\begin{table}[!ht]
\begin{adjustwidth}{-2.25in}{0in} 
\centering
\caption{
{\bf Example selected node pairs in decreasing order of the similarity.}}
\begin{tabular}{|l|l|l|l|l|l|l|l|}
\hline
node & node & similarity \\ \hline
2 &	3 &	0.4988\\
3 &	2 &	0.4988\\
5 &	10 &	0.3311\\
10 & 5 &	0.3311\\
1 &	2 &	0.2211\\
6 &	9 &	0.2209\\
9 &	5 &	0.2109\\
8 & 10 & 0.1667\\
4 &	8 &	0.1521\\
7 &	1 &	0.1456\\
\hline
\end{tabular}
\label{table1}
\end{adjustwidth}
\end{table}

{\bf Step 2: Community detection.} Communities are constructed by starting from the node pair with the maximum similarity and then including more node pairs from the list. In this example (Table~\ref{table1}), to begin with, node 2 and 3 form the first core-community $CCom_{1}=\{2,3\}$. Then node 5 and node 10 form the second core-community $CCom_{2}=\{5,10\}$. $CCom_{1}$ grows by including node 1 through its connection with node 2 and results in $CCom_{1}=\{2,3,1\}$. Then node 6 and node 9 form the third core-community $CCom_{3}=\{6,9\}$. Tide (9,5) connects $CCom_{2}$ and $CCom_{3}$. Node 8 is included into $CCom_{2}$ through its connections with node 10 results in $CCom_{2}=\{5,10,8\}$. Similarly, node 4 is added to $CCom_{2}$, node 7 is added to $CCom_{1}$.

$CCom_{1}=\{2,3,1,7\}.$

$CCom_{2}=\{5,10,8,4\}.$

$CCom_{3}=\{6,9\}.$

$Tide:\{9,5\}.$

$RCom_{1}=CCom_{1}=\{2,3,1,7\}.$

$RCom_{2}=\{CCom_{2},CCom_{3}\}=\{5,10,8,4,6,9\}.$

Tides are the selected node pairs that connect different core-communities. The tide (9,5) causes the merging of $CCom_{2}$ and $CCom_{3}$ into the real-community $RCom_{2}$. Core- and real- community correspond to two community detection level: diverse core level and small world real level. Diverse core level is the normal state of complex systems, while small world real level stands for the hidden inside community structure of complex systems.

Further detection could be done by regarding the detected communities as coarse-grained components. Then the coarse-grained system comprising these renormalized components can be further classified by steps 1 and 2.

\section*{Results And Analysis}

The citation data analyzed in this work are extracted from the CD version of Institute of Scientific Information journal citation report of SSCI dataset. There are 1575 journals have nonzero citation or cited contents, totally 837,001 citations. 

Without pre-given assumptions, the above MAX strategy detection generates the same number of 294 core- and real- communities, no tides. The results indicate that the communities at this resolution have weak intersections or interrelationships. This is because of the accurate measure of the similarity matrix. Detailed analysis of these 294 communities could be found in Ref.~\cite{29}. In order to explore the diversity of the SSCI journal system, two kinds of randomness are introduced in the selection of node pairs (step 1): rational random selection strategy (PSIM) and random selection strategy (P), where PSIM selects one node for every journal with a probability in proportion to similarity:

\begin{eqnarray}
\label{eq:schemeP}
p_{ij}=S_{ij}/(\sum_{k=1}^{N}S_{ik})
\end{eqnarray}

And P selects one node for each journal randomly. After selection, all the selected node pairs are sorted according to similarity in decreasing order, and community detection is carried out by step 2.

\begin{figure}[!h]
\centering
\renewcommand{\figurename}{Fig}
\caption{{\bf Evolution of the average number of core-community, real-community and tide at different probability of PSIM and P vs. MAX.} 
PSIM-CCom, PSIM-RCom and PSIM-tide are the average number of core-community, real-community and tide at different probability of rational random selection strategy PSIM; P-CCom, P-RCom and P-tide are the average number of core-community, real-community and tide at different probability of random selection strategy P.}
\label{fig1}
\end{figure}

Fig~\ref{fig1} is the evolution of the average number of two level communities and tides at different probability of PSIM and P strategy vs. MAX strategy. The number of core-communities and tides increase with the probability of randomness, whether the node pairs is selected by PSIM or P, while the number of real-communities is decreasing. The trend of the evolution shows that random P strategy brings diverse small world with less real-communities and more core-communities. And the rational randomness PSIM results in a more rational world with less core-communities and more real-communities than P. This is in accordance with the former research in complex networks that random generates small world network with no insider order~\cite{35}. However, the superiority of PSIM is not overwhelming. Further comparisons of PSIM and P are given by the detailed analysis of the community results below. 

\begin{figure}[!h]
\centering
\renewcommand{\figurename}{Fig}
\caption{{\bf Evolution of the average NMI the community results at different probability of PSIM or P vs. MAX.} (a) is the evolution of NMI at different probability of PSIM or P vs. MAX. PSIM-CCom and PSIM-RCom are the NMI of core- and real-community results at different probability of PSIM vs. MAX; P-CCom and P-RCom are the NMI of core- and real-community results at different probability of P vs. MAX. (b) is the evolution of NMI at different probability of PSIM vs. P. CCom is the NMI of the ore-community and RCom is the NMI of the real-community.}
\label{fig2}
\end{figure}

By setting the detected 294 communities with MAX strategy as standard community results, Fig~\ref{fig2} gives the evolution the average normalized mutual information (NMI) of the community detection results at different probability of PSIM and P vs. MAX. NMI is an efficient measure to evaluate community detection results~\cite{12}. It is defined as:

\begin{eqnarray}
\label{eq:schemeP}
NMI(X,Y)=\frac{H(X)+H(Y)-H(X,Y)}{(H(X)+H(Y))/2}
\end{eqnarray}

Where the 294 communities generated by MAX strategy are the standard structure Y, X is community structure detected with PSIM and P. H(x) and H(Y) are the entropy of community X and Y, H(X,Y) is the joint entropy of X and Y. 

Fig~\ref{fig2}(a) shows that, the NMI of core-communities decreases slowly and keep a high value above 0.7. And the NMI of real-community decreases fast to lower than 0.1. NMI of PSIM is always larger than P, it means that the communities generated by rational PSIM is always closer to the standard structure generated by MAX. However, rational random selection PSIM does not exhibit obvious superiority than simple random selection P strategy. And the closeness of PSIM-CCom and P-CCom shows that, the difference between the core-community structure for being rational or not is very small. More difference happens in the real-community structure. Fig~\ref{fig2}(b) is the evolution of the average normalized information of the community results at different probability of PSIM vs. P. The NMI of PSIM is $(0.7177-0.7057)/0.7057=1.7\%$ higher than P for the core-community, and $(0.1373-0.0803)/0.0803=70.98\%$ higher for the real-community. This confirms the fact that random connection generates small world local community with no inside order.

\begin{figure}[!h]
\centering
\renewcommand{\figurename}{Fig}
\caption{{\bf Evolution of NMI  with the increase of the number of available nth most similar journals.} 
CCom is the NMI of core-commuity, and RCom is the NMI of real-community.}
\label{fig3}
\end{figure}

Considering the slight superiority of PSIM in the above community detection process with complete similarity information We test two more strategies for the selection of node pairs. PSIM with limited number of nth most similar journals and MAX with different proportion of random deletion of similarity information. 

\begin{figure}[!h]
\centering
\renewcommand{\figurename}{Fig}
\caption{{\bf Evolution of NMI at different proportion of random deletion of similarity for each journal.} 
CCom is the NMI of core-commuity, and RCom is the NMI of real-community..}
\label{fig4}
\end{figure}

For the PSIM with limited nth most similar journals, Fig~\ref{fig3} is the evolution of NMI with the increase of the number of available journals in the decreasing order of similarity. The average NMI decrease quickly with the increase of available nth most similar journals. Incorporating with former results shown in Fig~\ref{fig2}, we can get the conclusion that it is not a good idea to detect community structure by PSIM strategy with too much information. For the 1575 SSCI journal system, the average NMI of real-community decreases below 0.5 after 30th available journals.

Finally, we test the efficiency of MAX selection with random deletion of similarity. Fig~\ref{fig4} is the evolution of average NMI with different proportion of random deletion of similarity for each journal. Fig~\ref{fig4} shows that MAX strategy is the only strategy which keeps the average NMI of real-community above 0.5, while the average NMI of core-community is bigger than 0.75 even $90\%$ of the similarity information are deleted.

\section*{Conclusion}

In this paper, we give a two level community detection for the SSCI journal system. Emerging characteristic numbers of core-community and real-community correspond to the two community detection level: diverse core level and small world real level. Diverse core level is the normal state of complex systems with big number of small communities, while small world real level stands for the hidden inside community structure of complex systems with small number of big communities. During the community detection process, we test five different strategy for the selection of node pairs. Comparison of the normalized mutual information show that, rational random selection (PSIM) is better than random selection (P) in finding stable community structure. However, rational random selection with too much information is almost the same with completely random selection. And maximum selection (MAX) is always the best strategy for community detection with complete or incomplete information. MAX strategy can not only overcome the resolution limit of modularity optimization~\cite{31},  but also detect uni-community. 

The success of maximum selection with incomplete information might indicates its efficiency in community detection for growing complex system. We will do further community detection for growing complex systems to demonstrate the efficiency of maximum selection with incomplete information. Advanced and detailed analysis of the evolution of tides will also be done in our future work to explore the overlapping property of community structure. 
\section*{Acknowledgments}

This work is supported by National Natural Science Foundation of China (Grant No. 11547003), and China Scholarship Council (Grant No. 201607620007). 


%
%
%

\end{document}